\documentstyle[12pt,epsf]{article}
\newcommand{\be}{\begin{equation}}
\newcommand{\ee}{\end{equation}}
\newcommand{\nn}{\nonumber}
\newcommand{\beba}{\begin{equation}\begin{array}{lcl}}
\newcommand{\eaee}{\end{array}\end{equation}}
\newcommand{\bea}{\begin{eqnarray}}
\newcommand{\eea}{\end{eqnarray}}
\newcommand{\ba}{\begin{array}}
\newcommand{\ea}{\end{array}}

\newcommand{\ns}{\normalsize}
\newcommand{\refs}[1]{(\ref{#1})}

\def\a{\alpha}

\def\g{\gamma}
\def\c{\chi}
\def\d{\delta}
\def\e{\epsilon}

\def\FO{\phi}

\def\k{\kappa}
\def\l{\lambda}
\def\m{\mu}
\def\n{\nu}
\def\o{\omega}
\def\p{\pi}

\def\r{\rho}
\def\s{\sigma}

\def\t{\tau}

\def\x{\xi}

\def\F{\Phi}
\def\G{\Gamma}
\def\J{\Psi}

\def\f{\Phi_0}

\def\ght{\hat{g}\kern-0.6em \widetilde{\raisebox{-0.12em}{\phantom{X}}}}
\def\bht{\hat{b}\kern-0.6em \widetilde{\raisebox{0.15em}{\phantom{X}}}}

\def\cL{{\cal L}}

\def\vf{\varphi}

\begin{document}

%%%%%%%%%%%%%%%%%%%%%%%%%%%%%%%%%%%%%%%%%%%%%%%%%%%%%%%%%%%%%%%%%%%%%%%%%%%

\begin{titlepage}
\title{\hfill{\ns UPR-774T, HUB-EP-97/50, IASSNS-HEP-97/106\\}
       \hfill{\ns hep-th/9709214\\[0.8cm]}
       {\Large\bf Soliton Solutions of M--theory on an Orbifold}}
\author{Zygmunt Lalak$^1\,$\setcounter{footnote}{3}\thanks{Supported in part
        by the Polish Commitee for Scientific Research grant 2 P03B 040 12}~,
        Andr\'e Lukas$^2\,$\setcounter{footnote}{0}\thanks{Supported in
        part by DOE under contract No.~DE-AC02-76-ER-03071}~~and
        Burt A.~Ovrut$^2\, ^3\, ^4\, {}^*$\\[0.5cm]
        {\ns $^1$Institute of Theoretical Physics, University of Warsaw} \\
        {\ns 00-681 Warsaw, Poland}\\[0.3cm]
        {\ns $^2$Department of Physics, University of Pennsylvania} \\
        {\ns Philadelphia, PA 19104--6396, USA}\\[0.3cm]
        {\ns $^3$Institut f\"ur Physik, Humboldt Universit\"at}\\
        {\ns Invalidenstra\ss{}e 110, 10115 Berlin, Germany}\\[0.3cm]
        {\ns $^4$School of Natural Sciences, Institute for Advanced Study}\\
        {\ns Olden Lane, Princeton, NJ 08540, USA}}
\date{}
\maketitle

\begin{abstract}
We explicitly construct soliton solutions in the low energy description of
M--theory on $S^1/Z_2$. It is shown that the 11--dimensional membrane
is a BPS solution of this theory if stretched between the $Z_2$
hyperplanes. A similar statement holds for the 11--dimensional 5--brane
oriented parallel to the hyperplanes. The parallel membrane and the
orthogonal 5--brane, though solutions, break all supersymmetries.
Furthermore, we construct the analog of the gauge 5--brane with gauge
instantons on the hyperplanes. This solution varies nontrivially along
the orbifold direction due to the gauge anomalies located on the
orbifold hyperplanes. Its zero mode part is identical to the weakly
coupled 10--dimensional gauge 5--brane.
\end{abstract}

\thispagestyle{empty}
\end{titlepage}

\section{Introduction}

One of the most interesting physical consequences of string duality is
the description of strongly coupled heterotic string theory as M--theory
on $S^1/Z_2$~\cite{hw1}. The low energy limit of this theory has been
constructed as 11--dimensional supergravity  coupled to two 10--dimensional
$E_8$ super--Yang--Mills theories on the two orbifold fixed
hyperplanes~\cite{hw2}. This construction allows one to study some of the
physics in the strongly coupled region of the heterotic string, despite the
fact that the fundamental underlying theory is still not fully known. More
precisely, the effective action of ref.~\cite{hw2} has been constructed
as an expansion in powers of the 11-dimensional Newton constant $\k$
where terms up to the first nontrivial order, that is $\k^{2/3}$ relative
to 11--dimensional supergravity, have been taken into account.

A wide class of vacuum solutions to the field-theoretical model of
\cite{hw1}, which  are relevant for a dimensional reduction down to $D=4$
theories with residual $N=1$
supersymmetry, has been constructed in ref.~\cite{w}. These solutions
correspond to Calabi--Yau 3--folds times $S^1/Z_2$ times 4--dimensional
Minkowski space in the zeroth order in $\k$. Once terms of order
$\k^{2/3}$ are taken into account, the Calabi--Yau space gets deformed due
to the nontrivial structure of the 3--form Bianchi identity.

In recent years, soliton solutions have played a crucial role 
in the study of dualities and nonperturbative effects in string and field 
theories. In particular, such solutions have been constructed for weakly
coupled heterotic string theory~\cite{str,symm,strom,dr}. If M--theory on
$S^1/Z_2$ indeed describes the strong coupling limit of heterotic string
theory, one expects to find the counterpart of those solutions as solitons of
M--theory on $S^1/Z_2$. Therefore, it is of interest to  
find the explicit soliton solutions of the field theoretical 
low energy description of M-theory on $S^1/Z_2$ constructed in \cite{hw1}.  

In this paper, we are going to construct several fundamental 
solutions for M-theory on $S^1/Z_2$, namely explicit solitonic solutions
that preserve a fraction of the supersymmetries of the original theory. 
Furthermore, we will compute these solitonic solutions to order $\k^{2/3}$,
thus including the non-trivial effects of the gauge and gravitational 
anomalies. Specifically, we will discuss the membrane, 5-brane and gauge
5-brane solitons of M-theory on $S^1 /Z_2$. Having these solutions in an
explicit form helps in verifying duality relations to other models.
Finally, it is interesting to study the implications of the nontrivial
boundary conditions at the hyperplanes on the amount of supersymmetry
supported by the solution. 

\vspace{0.4cm}

\section{General properties of the theory}

We start with a brief overview of M--theory on $S^1/Z_2$ which describes the
low energy limit of the strongly coupled heterotic string theory~\cite{hw2}.
The 11--dimensional coordinates are denoted by $x^0,...,x^9,x^{11}$. We take
$x^{11}$ as the orbifold direction and choose the range
$x^{11}\in [-\pi\r ,\pi\r ]$ with a periodic identification
$x^{11}\sim x^{11}+2\pi\r$ of the endpoints. The $Z_2$
symmetry acts as $x^{11}\rightarrow -x^{11}$ and, therefore, gives rise
to two 10--dimensional fixed hyperplanes at $x^{11}=0$ and $x^{11}=\pi\r$
respectively. The effective action for M--theory on $S^1/Z_2$ describes the
coupling of two 10--dimensional $E_8$ super--Yang--Mills theories on these
hyperplanes to 11--dimensional supergravity in the bulk. Let us denote
11--dimensional indices by $I,J,K,...=0,...,9,11$ and 10--dimensional
indices by $A,B,C,...=0,...,9$. Then the bosonic part of this action
is specified by
\be
 S = \int d^{11}x\left(\cL_{\rm SG}+\d (x^{11})\cL_{\rm YM}\right)\; ,\label{action}
\ee
with
\bea
 \cL_{\rm SG} &=& \frac{1}{2\k^2}\sqrt{-g}\left[ -R-\frac{1}{24}
                  G_{IJKL}G^{IJKL}\right]\nn\\
       && -\frac{1}{\k^2}\frac{\sqrt{2}}{3456}\e^{I_1...I_{11}}
          C_{I_1I_2I_3}G_{I_4...I_7}G_{I_8...I_{11}}+\; \cdots \label{LS}
\eea
and
\be
 \cL_{\rm YM} = -\frac{1}{4\l^2}\sqrt{-\,{}^{10}g}\; {\rm tr}(F_{AB}F^{AB})+\;
 \cdots \label{LB}
\ee
where $\l^2 = 2\pi (4\pi\k^2)^{2/3}$. Here, the dots indicate the omitted
fermionic terms which will be of no importance for the purpose of this
paper. $\cL_{\rm SG}$ is the usual Lagrangian of 11--dimensional
supergravity~\cite{D11}. The ``boundary'' Lagrangian $\cL_{\rm YM}$
describes an $E_8$ super--Yang--Mills theory at $x^{11}=0$ plus
additional (fermionic) terms which result from the coupling to
11--dimensional supergravity. The metric ${}^{10}g_{AB}$ is the restriction
of the 11--dimensional metric $g_{MN}$ to the $Z_2$ hyperplane. For
simplicity, we have concentrated on the hyperplane at $x^{11}=0$. The
contribution from the hyperplane at $x^{11}=\pi\r$
adds in an obvious way to the action~\refs{action} as well as to the following
formulae. The above action has to be supplemented with the nontrivial Bianchi
identity
\be
 (dG)_{11ABCD} = -3\sqrt{2}\frac{\k^2}{\l^2}\d (x^{11})\left(
                 {\rm tr}(F_{[AB}F_{CD]})-\frac{1}{2}
                 {\rm tr}(R_{[AB}R_{CD]})\right)\; ,
 \label{Bianchi}
\ee
for the 4--form field strength $G_{IJKL}$. It has been derived in
ref.~\cite{hw2} from the requirements of anomaly cancellation and
supersymmetry and it is designed to reproduce the analogous equation for
the heterotic string upon reduction to 10 dimensions. In particular, the
factor of $1/2$ in front of the tr$(R^2)$ serves to distribute the
total gravitational contribution equally to the two hyperplanes.
The Bianchi identity~\refs{Bianchi} can be solved in terms of the 3--form
field $C_{IJK}$ and the gravity and Yang--Mills Chern--Simons forms
$\o^{(L)}_{ABC}$, $\o_{ABC}$ as follows
\bea
 G_{ABCD} &=& \partial_AC_{BCD}\pm 23\;{\rm perm.} \\
 G_{11BCD} &=& (\partial_{11}C_{BCD}\pm 23\;{\rm perm.})+\frac{\k^2}
               {\sqrt{2}\l^2}\d (x^{11})(\o_{BCD}-\frac{1}{2}\o^{(L)}_{BCD})
               \; .
\eea
Explicitly, the Yang--Mills Chern--Simons form is given in terms of the
gauge field as
\be
 \o_{ABC} = {\rm tr}\left[A_A(\partial_BA_C-\partial_CA_B)+\frac{2}{3}A_A
            [A_B,A_C]+\; \mbox{cyclic perm.}\right]\; .
\ee
A similar expression holds for $\o^{(L)}_{ABC}$. 
Let us now collect the bosonic equations of motion to be derived from
the action~\refs{action} which we are going to need for our discussion of
soliton solutions. For the explicit examples, we will find that the
gravitational contribution to the anomaly in eq.~\refs{Bianchi} vanishes.
Consequently, we drop the resulting terms from the following equations of
motion. The Einstein equation is given by
\be
 R_{MN}-\frac{1}{2}g_{MN}R+\frac{1}{6}\left( G_{MJKL}{G_N}^{JKL}-\frac{1}{8}
      g_{MN}G_{IJKL}G^{IJKL}\right)+\d (x^{11})T_{MN} = 0\; ,
 \label{Einst}
\ee
where the only nonvanishing components of the Yang--Mills stress energy
tensor $T_{MN}$ are
\be
 T_{AB} = \frac{\k^2}{\l^2} (g_{11,11})^{-\frac{1}{2}}\left[ 
          {\rm tr}(F_{AC}{F_B}^C)
          -\frac{1}{4}g_{AB}{\rm tr}(F_{CD}F^{CD})\right]\; .
\ee
The 3--form equations of motion read
\bea
 \partial_M\left(\sqrt{-g}G^{MNPQ}\right)&-&\frac{\sqrt{2}}{1152}
 \e^{NPQI_4...I_{11}}\, G_{I_4...I_7}G_{I_8...I_{11}}\nn\\
 &+&\frac{\k^2}{\l^2}\frac{1}{432}\d (x^{11})\e^{11NPQA_5...A_{11}}
 \o_{A_5A_6A_7}G_{A_8...A_{11}} = 0\; .\label{eomG}
\eea
Finally, for the gauge fields we have
\be
 D_AF^{iAB}-\frac{1}{\sqrt{2}} (g_{11,11})^{\frac{1}{2}} F_{DE}^i 
G^{11BDE}
 -\frac{1}{1152}\frac{1}{\sqrt{-\,{}^{10}g}}\, \e^{BB_2...B_{10}}A^i_{B_2}
    G_{B_3...B_6}G_{B_7...B_{10}}=0\; , \label{eomF}
\ee
with the space--time and gauge covariant derivative $D_A$. It is
important to restrict the solutions of the above set of equations to those
which respect the $Z_2$ orbifold symmetry. This $Z_2$ invariance implies for
the fields
\bea
  g_{AB}(x^{11}) = g_{AB}(-x^{11})&\quad& G_{ABCD}(x^{11})=-G_{ABCD}(-x^{11})
  \nn \\
 g_{A11}(x^{11}) = -g_{A11}(-x^{11})&\quad& G_{11BCD}(x^{11})=
 G_{11BCD}(-x^{11}) \label{Z2}\\
 g_{11,11}(x^{11}) = g_{11,11}(-x^{11}&\quad&C_{ABC}(x^{11}) =
  -C_{ABC}(-x^{11})\nn \\
 &\quad&C_{11BC}(x^{11})=C_{11BC}(-x^{11})\nn
\eea
Clearly, there is no condition on the gauge fields, since they are defined on
the $Z_2$ hyperplanes only. Furthermore, to check the number of preserved
supersymmetries for the solutions we are going to consider, we will need the
supersymmetry transformations of the gravitino $\J_M$ and the gauginos $\c^\a$
\bea
 \d\J_M &=& D_M\eta +\frac{\sqrt{2}}{288}\left({\G_M}^{IJKL}-8\d_M^I\G^{JKL}
          \right)\eta G_{IJKL}+\; \cdots \label{susy}\\
 \d \c^\a &=& -\frac{1}{4}\G^{AB}F_{AB}^\a\eta +\; \cdots\; .
\eea
The dots denote terms that involve the fermion fields of the
theory. These terms vanish for the purely bosonic solution we are interested
in. The 11--dimensional gamma matrices obey the condition
$\{\G_M,\G_N\} =2g_{MN}$. In order to keep the transformation~\refs{susy}
compatible with the $Z_2$ symmetry, the 11--dimensional Majorana spinor
$\eta$ has to be restricted by
\be
 \eta (x^{11}) = \G_{11}\eta (-x^{11})\; .\label{Z2eta}
\ee
Note that this condition by itself does not restrict the number of 
11--dimensional supersymmetries. On the $Z_2$ hyperplane, however,
we have the chirality condition $\eta (0) = \G_{11}\eta (0)$ which leads
to the correct amount of supersymmetry, $N=1$, in 10 dimensions.

\vspace{0.4cm}

\section{The membrane}

We are now ready to discuss BPS solutions of the above theory. To this end,
it is useful to observe that the Yang--Mills boundary theory $\cL_{\rm YM}$ in
eq.~\refs{action}, as well as the nontrivial term in the Bianchi
identity~\refs{Bianchi}, are suppressed by $\k^{2/3}$ with respect to
the bulk theory $\cL_{\rm SG}$. To lowest order in $\k$, the theory can
thus be viewed as 11--dimensional supergravity subject to the $Z_2$
constraints~\refs{Z2}. One approach to finding BPS solutions
is, therefore, to start with such a solution of 11--dimensional
supergravity; that is, with the elementary BPS membrane or the solitonic BPS
5--brane, and analyze to what extent it generalizes to a BPS solution
of M--theory on $S^1/Z_2$. This requires a discussion of $Z_2$ invariance
as well as $\k^{2/3}$ corrections. We will follow this approach for
the $D=11$ membrane and 5--brane.

\vspace{0.4cm}

Let us briefly review the multi--membrane solution of 11--dimensional
supergravity~\cite{ds,dr}. It is specified by the Ansatz~\cite{ds,dr}
\bea
 ds^2 &=& e^{2A}dx^\m dx^\n\eta_{\m\n}+e^{2B}dy^mdy^n\d_{mn} \nn \\
 C_{\m\n\r} &=& \pm\frac{1}{6\sqrt{2}\;{}^3g}\e_{\m\n\r}e^C \label{m_ans}\\
 G_{m\m\n\r} &=& \pm\frac{1}{\sqrt{2}\;{}^3g}\e_{\m\n\r}\partial_m e^C\; .\nn
\eea
Here $x^\m$ are the $2+1$ worldvolume coordinates labeled by indices
$\m ,\n ,...$ and $y^m$ are the $8$ transverse coordinates labeled by
indices $m,n,...$. Furthermore, ${}^3g$ is the determinant of the
worldvolume part of the metric. The functions $A$, $B$, $C$ depend on the
transverse coordinates $y^m$ only. This Ansatz represents a multi--membrane
solution of 11--dimensional supergravity (strictly, coupled to the
11--dimensional supermembrane action) provided that $A=C/3$ and $B=-C/6$. Here
$e^{-C}$ should be a harmonic function; that is, it should fulfill
$\Box e^{-C}=0$ away from the singularities where
$\Box\equiv \d^{mn}\partial_m\partial_n$ is the transverse Laplacian.
For a solution corresponding to membrane sources
at ${\bf y}^{(i)}=(y^{(i)m})$, 
the harmonic function $e^{-C}$ can be written as
\be
 e^{-C} = 1+\sum_i\frac{1}{|{\bf y}-{\bf y}^{(i)}|^6}\; .
 \label{m_harm}
\ee
With the above relations between $A$, $B$ and $C$, the supersymmetry
variation of the gravitino in eq.~\refs{susy} vanishes for spinors
$\eta$ of the form
\be
 \eta = \e\otimes\r\; ,\quad \r =\r_0\; e^{C/6}\; ,\label{m_eta}
\ee
with constant 3-- and 8--dimensional spinors $\e$, $\r_0$ and $\r_0$
satisfying the chirality condition
\be
(1\pm\s )\r_0 = 0\; .\label{m_chir}
\ee
In these formulae, the 11--dimensional gamma matrices have been split as
$\G_M=\{ \g_\m\otimes\s ,1\otimes\s_m\}$, where $\g_\m$, $\s_m$ are the
3-- and 8--dimensional gamma matrices, respectively, and $\s = \prod_m\s_m$.
The projection condition~\refs{m_chir} states that the solution preserves
$1/2$ of the original $D=11$ supersymmetry. The sign in eq.~\refs{m_chir}
which determines the chirality of the preserved supersymmetry is the same
as in the Ansatz~\refs{m_ans}.

Next, we would like to embed this solution into M--theory on $S^1/Z_2$.
This requires, as a first step, a discussion of its $Z_2$ properties.
There are two different ways to orient the membranes with respect to the
$x^{11}$--direction, namely to choose $x^{11}$ as a worldvolume or a
transverse coordinate. In the first case, the membranes stretch between
the two $Z_2$ hyperplanes and intersect them as $1+1$--dimensional
extended objects; that is, as strings. In the second case, the membranes
are parallel to the $Z_2$ hyperplanes. Let us first assume that
$x^{11}$ is a worldvolume direction. In this case, there is no explicit
dependence on $x^{11}$ in the solution~\refs{m_ans}, so that the only
nonvanishing fields should be the $Z_2$--even ones. A comparison with the
$Z_2$ conditions~\refs{Z2} shows that indeed all fields in the
eq.~\refs{m_ans} are $Z_2$ even. The ``orthogonal'' membrane therefore
automatically satisfies the orbifold constraint. In addition, to find the
number of preserved supersymmetries, we have to implement the $Z_2$
constraint~\refs{Z2eta} on the spinor $\eta$. With
$\G_{11} = \g_{11}\otimes\s$, eqs.~\refs{Z2eta}, \refs{m_chir}
imply
\be
 (1\pm\s )\r_0 = 0\; ,\quad (1\pm\g_{11})\e = 0\; ,\label{m_chir1}
\ee
for $\eta$ as defined in eq.~\refs{m_eta}. Again, the sign
which determines the chirality of the preserved supersymmetry is the same
as in the Ansatz~\refs{m_ans}. These conditions show that the
membrane solution preserves $1/4$ of the 11--dimensional supersymmetry
of M--theory on $S^1/Z_2$ and $1/2$ of the supersymmetry on the
10--dimensional hyperplanes. It is in this sense that we will use the
term ``BPS state of M--theory on $S^1/Z_2$'' in the following.
So far we have only considered terms to lowest order in $\k$. How is the
solution affected by the corrections of order $\k^{2/3}$?
Since we have not turned on gauge fields, the only source of such a correction
is the tr$(R^2)$ term in the Bianchi identity~\refs{Bianchi}. It is, however,
straightforward to show that tr$(R^2)$ vanishes for the Ansatz~\refs{m_ans}
and, consequently, the solution does not receive any corrections of order
$\k^{2/3}$.
To summarize, we have therefore seen that the BPS membrane solution
of $D=11$ supergravity stretched between the $Z_2$ hyperplanes is also
a BPS solution of M--theory on $S^1/Z_2$, including corrections of relative
order $\k^{2/3}$. Upon restriction to the hyperplanes, the solution reduces
to a string solution in the same way the membrane of $D=11$
supergravity reduces to a string by dimensional reduction of one of its
worldvolume coordinates~\cite{ds}. 
This result extends immediately to 
multi--membrane solutions ending in multi--strings on the boundary 
hyperplanes.

It remains to show that the singularities in the above solution arise
from supermembrane source terms~\cite{ds}. For this to be the case,
the source terms and, hence, the supermembrane equations of motion, must be
compatible with the $Z_2$ orbifold symmetry. The gauge anomaly of the 
supermembrane worldvolume action embedded into a target space manifold with
boundaries has been discussed in ref.~\cite{ceder}. 
Since, for the membrane solution, the gauge and gravitational anomalies
are not switched on, we are allowed to consider the simple supermembrane
action without anomaly cancellation terms for our purpose. The
bosonic part of this action is
\be
S_{\rm SM} = \int d^3 \x \left( -\frac{1}{2} \sqrt{-\gamma} \gamma^{ij}
\partial_i X^M \partial_j X^N g_{MN} +
\frac{1}{2}\sqrt{-\gamma} \pm  \frac{1}{3!} \e^{ijk} \partial_i X^M
\partial_j X^N  \partial_k X^P C_{MNP}\right)
\label{eq:wv}
\ee
where $\x^i,\; i=0,1,2$ are the worldvolume coordinates, $\gamma_{ij}$
is the worldvolume metric and $X^M$ are the 11--dimensional target space 
coordinates. It follows from the identification of $X^M$ with 
$x^M$ in the equations of motion for $S+S_{\rm SM}$ that the target space
coordinates should transform as $X^A\rightarrow X^A$,
$X^{11} \rightarrow -X^{11}$ under the $Z_2$ symmetry. Furthermore, this
identification tells us that the background fields $g_{MN}(X^Q)$ and
$C_{MNP}(X^Q)$ should satisfy the $Z_2$ conditions~\refs{Z2} with
$x^M$ replaced by $X^M$. Though the following discussion can be carried
out in general, it is enough for our purpose to consider the specific
gauge
\be 
X^{\m} = \x^{\m}\; \quad \m=0,1,11\; ,
\label{eq:gauge}
\ee
which is adapted to the orientation of the membrane worldvolume parallel
to the orbifold direction. Clearly, this leads to the $Z_2$
transformation $\x^i\rightarrow \x^i$, $i=0,1$ and
$\x^{11}\rightarrow -\x^{11}$ for the worldvolume coordinates.
Then it is straightforward to show that the membrane equations
of motion are $Z_2$ covariant provided we require for the worldvolume
metric that
\bea
& \gamma_{ij}(\x^{11}) = \gamma_{ij} (-\x^{11})&
   \nn \\
& \gamma_{i11}(\x^{11}) =-\gamma_{i11}(-\x^{11})&
   \nn \\
& \gamma_{11,11}(\x^{11}) = \gamma_{11,11}(-\x^{11})&
\label{eq:mwv}
\eea
where $i,j = 0,1$. This follows easily from the $\g$--equation of motion,
$\g_{ij} =\partial_iX^M\partial_jX^Ng_{MN}$, using the above gauge choice and
the $Z_2$ properties of the metric. We conclude that, for an appropriate
extension of the $Z_2$ symmetry to the worldvolume coordinates and a
restriction of the worldvolume metric as above, the 
supermembrane equations of motion are $Z_2$ covariant. The explicit solution
which we need to support the singularity is, in the 
$X^\m = \x^\m$, $\m = 0,1,11$ gauge, $X^m = \mbox{const}$, $m=3,...,9$ and
$\g_{\m\n} = e^{2A}\eta_{\m\n}$. According to the above rules, this solution
indeed respects the $Z_2$ symmetry and, therefore, provides an acceptable
source term for the ``parallel'' membrane solution of M--theory on $S^1/Z_2$.
In addition, we should check that this solution does not break any of the
preserved supersymmetries. This can be done in exactly the same way as
for the ordinary 11--dimensional membrane~\cite{ds} and is guaranteed by
the condition~\refs{m_chir}.

Let us now address the case of $x^{11}$ as a transverse direction. First,
we should guarantee the $Z_2$ invariance of the harmonic function $e^{-C}$.
This is easily done by pairing each membrane source at $y^{(i)11}$ with
a ``mirror source'' at $-y^{(i)11}$ in the expression~\refs{m_harm}. Then
all metric components in eq.~\refs{m_ans} are $Z_2$ invariant.
The components $G_{11\m\n\r}$ of the 4--form in eq.~\refs{m_ans}, however,
are proportional
to $\partial_{11}e^C$ which changes sign under $x^{11}\rightarrow -x^{11}$.
This is in conflict with the $Z_2$ conditions~\refs{Z2}. One can
cure this problem by using the additional sign freedom in eq.~\refs{m_ans};
that is, by choosing the $+$ sign for $x^{11}\in [0,\pi\r ]$ and the $-$
sign for $x^{11}\in [-\pi\r,0 ]$. Previously we had the chirality
conditions~\refs{m_chir1} either for the positive or the negative sign.
Now the conditions~\refs{m_chir1} have to be simultaneously fulfilled for
both signs, so that all components of the spinor are projected out.
Therefore, though this is a way of constructing
a solution of M--theory on $S^1/Z_2$ based on the 11--dimensional membrane
parallel to the hyperplanes, this solution does not respect any of the
supersymmetries of M--theory on $S^1/Z_2$. Clearly, by the same argument
as previously, the solution receives no corrections from $\k^{2/3}$ terms.

\vspace{0.4cm}

\section{The 5--brane}

Next, we will carry out a similar discussion for the 5--brane of $D=11$ 
supergravity~\cite{gue}. The Ansatz for this solution is given by
\bea
 ds^2 &=& e^{2A}dx^\m dx^\n\eta_{\m\n}+e^{2B}dy^mdy^n\d_{mn} \nn \\
 G_{mnrs} &=& \pm\frac{1}{\sqrt{2}}e^{-8B}{\e_{mnrs}}^t\partial_te^C\; ,
 \label{5_ans}
\eea
where $\m ,\n ,...$ label time and the 5 spatial worldvolume directions
and $m,n,...$ the 5 transverse directions. This Ansatz solves the equations
of motion of 11--dimensional supergravity provided $A=-C/6$, $B=C/3$ and
$e^{-2C}\Box e^C = 0$, where $\Box = \d^{mn}\partial_m\partial_n$. For
5--branes at ${\bf y}^{(i)}=(y^{(i)m})$ the harmonic function $e^C$ can be
written in the form
\be
 e^C = 1+\sum_i\frac{1}{|{\bf y}-{\bf y}^{(i)}|^3}\; .\label{5_harm}
\ee
With the above relations between $A$, $B$ and $C$, the gravitino supersymmetry
variation~\refs{susy} vanishes for spinors
\be
 \eta=\e\otimes\r\; ,\quad \r = e^{-C/12}\r_0
\ee
with
\be
 (1\pm\g )\e = 0\; .\label{5_chir}
\ee
Here, $\e$, $\r_0$ are constant 6-- and 5--dimensional spinors, respectively.
The sign in eq.~\refs{5_chir} which determines the chirality of the unbroken
supersymmetries is the same as the one in the Ansatz~\refs{5_ans}. We have
decomposed the $D=11$ gamma matrices as $\G_m = \{\g_\m\otimes 1,
\g\otimes\s_m\}$ with $\g = \prod_\m \g_\m$. As before, to discuss $Z_2$
invariance, we distinguish the two cases of $x^{11}$ being a worldvolume
or a transverse direction. Let us start with the latter case. $Z_2$ invariance
of the harmonic function $e^C$ is achieved by pairing each 5--brane at
$y^{(i)11}$ with a mirror 5--brane at $-y^{(i)11}$ in 
eq.~\refs{5_harm}. Comparison
with the $Z_2$ condition~\refs{Z2} shows that this guarantees a $Z_2$--even
solution. Using $\G_{11}=\g\otimes\s_{11}$ and the eqs.~\refs{Z2eta},
\refs{5_chir} we get
\be
 (1\pm\g )\e = 0 \; ,\quad (1\pm\s_{11})\r_0 = 0\; . \label{5_chir1}
\ee
Therefore $1/4$ of the 11-dimensional supersymmetry and $1/2$ of the
10--dimensional supersymmetry of M--theory on $S^1/Z_2$ is
preserved. As in the membrane, tr$(R^2)$ vanishes for the metric~\refs{5_ans}
so that the ``parallel'' 5--brane receives no corrections of order $\k^{2/3}$.
In summary, we have seen that the 11--dimensional 5--brane oriented parallel
to the $Z_2$ hyperplanes is a BPS solution (understood in the sense explained
above) of M--theory on $S^1/Z_2$ including terms of order $\k^{2/3}$.

For the 5--brane stretched between the hyperplanes, we face a similar
situation as for the ``parallel'' membrane. In this case, the function
$e^C$ is automatically $Z_2$ invariant. The components $G_{mnrs}$ in
eq.~\refs{5_ans}, however, do not change sign at the hyperplanes as required
by $Z_2$ invariance unless we use the sign freedom in the
Ansatz~\refs{5_ans}. As before, this means that we choose the $+$ sign for
$x^{11}\in [0,\pi\r ]$ and the $-$ sign for $x^{11}\in [-\pi\r,0 ]$.
This leads to a solution of M--theory on $S^1/Z_2$
including $\k^{2/3}$ terms. However, since the sign in the Ansatz~\refs{5_ans}
is linked to the sign in the chirality condition~\refs{5_chir},
eq.~\refs{5_chir1} holds for both signs simultaneously. Consequently, no
supersymmetries are preserved.

\section{The gauge 5--brane}

The solutions derived from 11--dimensional supergravity which we have
considered so far did not lead to any  nontrivial term on the right hand side
of the Bianchi identity~\refs{Bianchi}, since tr$(R^2)=0$. 
A way to obtain a 
nontrivial  Bianchi identity, is to consider a solution with a nonvanishing
gauge field configuration, so that tr$(F^2)\neq 0$. For the weakly coupled
heterotic string, such a solution is given by the gauge 5--brane of
ref.~\cite{strom}. We are now going to analyze the analog of this solution
in the strongly coupled case.

Since tr$(R^2)=0$ for the metric which we are going to consider, the Bianchi
identity~\refs{Bianchi} now reads
\be
 (dG)_{11ABCD} = -3\sqrt{2}\, \frac{\k^2}{\l^2}\left( \d (y^{11})
                 \mbox{tr}(F^{(1)}_{[AB} F^{(1)}_{CD]})+
                 \d (y^{11}-\p\r )\mbox{tr}(F^{(2)}_{[AB} F^{(2)}_{CD]})
                 \right)\; ,
 \label{Bianchi1}
\ee
where we have included gauge fields strengths $F^{(1)}$ and $F^{(2)}$ for
both hyperplanes. Later on, we will find it useful to solve this Bianchi
identity, as well as the equation of motion for $G$,
 in the boundary picture~\cite{hw2}
as opposed to the orbifold picture which we have used so far. In this
picture, we think of the 11--dimensional space as the interval
$0\leq x^{11}\leq\pi\r$ times a 10--dimensional manifold. Then the source terms
in the orbifold picture turn into boundary conditions at the two boundaries
$x^{11}=0,\pi\r$ of the 11--dimensional manifold. More explicitly, one can
determine these boundary conditions by solving the above Bianchi identity
close to the fixed points~\cite{hw2}. The resulting conditions are
\bea
 \left. G_{ABCD}\right|_{x^{11}=0} =  -\frac{3}{\sqrt{2}}
 \frac{\k^2}{\l^2} \mbox{tr}(F^{(1)}_{[AB}F^{(1)}_{CD]})\nn\\
 \left. G_{ABCD}\right|_{x^{11}=\pi\r} =  +\frac{3}{\sqrt{2}}
 \frac{\k^2}{\l^2} \mbox{tr}(F^{(2)}_{[AB}F^{(2)}_{CD]})\; .
 \label{Bianchi3}
\eea
One then  solves the homogeneous Bianchi
identity $dG=0$ instead of eq.~\refs{Bianchi1}, with the equation of
motion~\refs{eomG} for $G$ being subject to the conditions~\refs{Bianchi3}.

Let us now set up the Ansatz for the gauge 5--brane solutions.
The previous experience with the ordinary 5--brane leads us to orient the
$x^{11}$--direction in the transverse space in order 
to preserve some supersymmetries.
We therefore start with the Ansatz
\be
 ds^2 = e^{2A}dx^\m dx^\n\eta_{\m\n}+e^{2B}dy^mdy^n\d_{mn}
        \label{g5_metr}
\ee
for the metric, where $\m ,\n , ...=0,...,5$ label the worldvolume directions
and $m,n,...=6,...,9,11$ the transverse directions including $x^{11}$.
We also introduce indices $a,b,...=6,...,9$ for the transverse directions
orthogonal to the orbifold. For the 4--form we write, in analogy with the
ordinary 5--brane,
\be
 G_{mnrs} = \pm\frac{1}{\sqrt{2}}e^{-8B}{\e_{mnrs}}^t\partial_t e^C\; .
 \label{5g_ans}
\ee
Finally, we have to specify the gauge fields. We consider simple $SU(2)$
instantons~\cite{inst} on both $Z_2$ hyperplanes specified
by $A^{(1,2)i}_a =\eta^{(1,2)i}_{ab}\partial_b h_{1,2}$ where
$h_{1,2}=-\ln f_{1,2}$. Here the tensors $\eta^{(1,2)i}_{ab}$ are defined by
$\eta^{(1,2)i}_{ab}=\e_{0iab}+\e^{(1,2)}(\d_{ia}\d_{b0}-\d_{ib}\d_{a 0})$
and the function $f$ satisfies $f^{-1}\Box_4 f = 0$ with the 10--dimensional
transverse Laplacian $\Box_4\equiv\d^{ab}\partial_a\partial_b$.
The signs $\e^{(1,2)}=\pm$ in the definition of $\eta^{(1,2)i}_{ab}$ 
specify whether the instanton is selfdual ($+$ sign) or anti--selfdual
($-$ sign). The $SU(2)$ generators are chosen as $T^i = \t^i /2i$ with
the Pauli matrices $\t^i$. For the gauge fields defined in such a way,
one finds $\mbox{tr}(F_{[ab}^{(1,2)} F_{cd]}^{(1,2)}) =\pm\frac{1}{6}
\hat{\e}_{abcd}\Box_4^2 h_{1,2}$ where $\hat{\e}_{abcd}$ is the flat
totally antisymmetric 
$\e$ tensor. Thus, the right hand side of the Bianchi identity~\refs{Bianchi1}
or, equivalently, the boundary conditions \refs{Bianchi3} are completely
determined.

We are now ready to discuss the equations of motion. Since the Ans\"atze
for the metric and the 4--form are identical to the ones for the ordinary
5--brane, it is natural to look for a solution which fulfills the
familiar relations $A=-C/6$, $B=C/3$. Indeed, using these relations one
finds that the $(mn)$ components of the Einstein equation~\refs{Einst},
the equation of motion~\refs{eomG} for $G$ and the equations of motion
for the gauge fields~\refs{eomF} are identically fulfilled.
The remaining equations are the $(\m \n )$ components of the Einstein
equation and the Bianchi identity~\refs{Bianchi1} which both contain
gauge field source terms. If we choose the instantons on the hyperplanes
to be of the same type (both selfdual or both anti--selfdual) and, in
addition, choose (anti)--selfdual instantons for the $+$ ($-$) sign in
the Ansatz~\refs{5g_ans} for $G$ ($\e^{(1)}=\e^{(2)}=\pm$), then these
two equations are, in fact, identical to
\be
 \Box_4 e^C = \d(y^{11})J_1+\d (y^{11}-\pi\r )J_2\; .\label{box50}
\ee
Here $\Box_5\equiv\d^{mn}\partial_m\partial_n$ and $J_1$, $J_2$ are
the instanton source terms explicitly given by
\be
 J_{1,2} = -\frac{\k^2}{2\l^2}\Box_4^2h_{1,2}
\label{eq:curr}
\ee
with the functions $h_{1,2}$ as defined above. The remaining problem
is to solve equation~\refs{box50}, and we will find it useful to do this
in the boundary picture. Following the steps explained in the beginning
of this section, the equivalent problem in the boundary picture can
be formulated as
\be
\Box_5 \Phi (x^a; x^{11}) =0
\label{eq:box5}
\ee
with boundary conditions 
\be
\partial_{11} \Phi |_{x^{11}=0} = \frac{J_1(x^a)}{2}, \;
\partial_{11} \Phi |_{x^{11}=\pi\r} = - \frac{J_2(x^a)}{2}\label{bound}
\ee
Here we have defined $\Phi = e^C$. In this formulation, it is obvious that
the solution will have a nontrivial dependence on the $x^{11}$--coordinate
since the field $\Phi$ has to interpolate between the ``surface charges''
on the boundaries provided by the instantons. More specifically,
since M--theory on $S^1/Z_2$ turns into weakly coupled heterotic string
theory upon shrinking the $x^{11}$ direction, we expect an
$x^{11}$--independent bulk component of $\Phi$ which, in some sense (to be
precisely specified later), corresponds to the weakly coupled gauge 5--brane.
On top of this bulk component, $\Phi$ contains an $x^{11}$--dependent piece
which represents the strong coupling regime  ``dressing'' of 
the gauge 5--brane.
In general, this solution for $\Phi$ cannot be expressed in terms of the
instanton sources $J_{1,2}$ in a simple way. Noticing that eqs.~\refs{eq:box5},
\refs{bound} constitute a problem in potential theory  with von Neumann 
boundary
conditions,
 one might apply methods familiar from classical electrodynamics
to find a solution. Here, we will, however, use a more direct 
approach which is better suited to the expected structure of the solution.
Let us split $\Phi$ into two parts as
\be
 \Phi = \Phi_0+\phi, \label{ansa}
\ee
where $\phi$ is a function of $x^m, \; m=6,7,8,9,11$,
and $\Phi_0$ is a function of $x^a, \; a=6,7,8,9$ only. The average $<...>$ 
over
the 11-th dimension is defined as
$<f> = \frac{1}{\pi\r} \int_0^{\pi\r} f(x^{11}) d x^{11}$. 
We further demand that 
\be
<\phi> = 0.
\ee
The 
condition $<\phi> = 0$ can always be achieved by a redefinition of $\Phi_0$,
and it determines the decomposition~\refs{ansa} uniquely.  The idea is that
the $x^{11}$--independent piece $\Phi_0$ is the zero mode of $\Phi$ and will
eventually correspond to the 10--dimensional gauge 5--brane,
whereas $\phi$ represents the $x^{11}$--dependent corrections. Inserting
this Ansatz into eq.~\refs{eq:box5} one gets
\be
\Box_5 \F = \Box_4 \f + \Box_4 \FO + \partial_{11}^2 \FO =0.
\label{eq:BB}
\ee
In addition, it follows from eq.~\refs{bound} that $\phi$ has to fulfill
the boundary condition
\be
\partial_{11} \phi |_{x^{11}=0} = \frac{J_1(x^a)}{2}, \;\quad
\partial_{11} \phi |_{x^{11}=\pi\r} = - \frac{J_2(x^a)}{2}\; .\label{bound0}
\ee
By integrating over $x^{11}$, taking into account $<\phi >=0$ and
eq.~\refs{bound0}, we arrive at a purely 4--dimensional equation for $\Phi_0$
\be
\Box_4 \f = \frac{1}{2 \pi\r} (J_1 + J_2) \; .\label{B4}
\ee
On the other hand, using this result to eliminate $\Phi_0$ from
eq.~\refs{eq:BB} we find $\phi$ to be determined by
\be
\Box_4 \FO +  \partial_{11}^2 \FO = - \frac{1}{2 \pi\r} (J_1 + J_2)\; .
\label{eq:fo}
\ee
What we have achieved so far is to split the original equation into an
$x^{11}$--independent equation~\refs{B4} for $\Phi_0$, and a modified
boundary value problem for $\phi$. Together, the equations~\refs{B4},
\refs{eq:fo} and~\refs{bound0} are completely equivalent to the original
problem. If we put $J_2=0$, eq.~\refs{B4} is just
the same equation that arises for the weakly coupled gauge
5--brane. Its solutions are therefore well known~\cite{strom}. 
Consider the case of two instantons, one located on the $x^{11}=0$
hyperplane at $r=0$ (where $r\equiv\sqrt{y^ay^b\d_{ab}}$ is the 
4--dimensional radius) with size specified by 
$\sigma_1$  and $h_1=-\ln (1+\frac{\s_1^2}{r^2})$,  and the other 
on the $x^{11} = \pi \rho$ hyperplane, also at $r=0$, with size $\sigma_2$ and 
$h_2=-\ln (1+\frac{\sigma_2^2}{r^2})$. Then the 
nonsingular solution of equation \refs{B4} is given by
\be
\f = 1 + \frac{2 \k^2}{\pi\r \lambda^2} \left (
\frac{2 \sigma_1^2+ r^2}{(\sigma_1^2 + r^2)^2} +
\frac{2 \sigma_2^2+ r^2}{(\sigma_2^2 + r^2)^2} \right )
\; . 
\label{eq:sols}
\ee 
The arbitrary additive constant in $\Phi_0$ has been normalized to $1$,
so that
the physical distance between the two hyperplanes 
far away from the instanton core
($r\rightarrow\infty$) equals the coordinate distance $\pi\r$.
Note that allowing formally $\sigma_2 \rightarrow 0$, and subtracting 
$ \frac{2 \k^2}{\pi\r \lambda^2 r^2}$ which is a solution to the homogeneous 
equation, corresponds 
to the special case of an instanton at $x^{11}=0$ but no instanton 
on the $x^{11}=\pi \rho$
hyperplane. In this case, the above solution becomes
\be
\f = 1 + \frac{2 \k^2}{\pi\r \lambda^2}
\frac{2 \sigma^2+ r^2}{(\sigma^2 + r^2)^2}
\; . 
\label{eq:sols2}
\ee 
which corresponds to Strominger's original solution. 
The generalization of this solution to two instantons located at different 
positions, as well as to
the multi--instanton case, is
straightforward.

The final task is to find the solution for $\phi$ which contains the
nontrivial dependence on $x^{11}$. In general, the solution of eq.~\refs{eq:fo}
is very complicated because of the mixing of $x^{11}$ with the 4--dimensional
coordinates. A solution can, however, be obtained by considering the following
expansion 
\be
\FO = \sum_{n=0}^{n=\infty} \psi_n\label{series}
\ee
where 
\be 
\psi_n = P_n (x^{11}) \Box^n_4 J_1 + Q_n (x^{11}) \Box^n_4 J_2\; .
\label{eq:psin}
\ee
Here, $P_n$ and $Q_n$ are functions of $x^{11}$. Inserting this series
into eq.~\refs{eq:fo}, and applying the boundary conditions~\refs{bound0} as
well as $<\phi>=0$, we arrive at the recurrent series of equations
\bea
\partial_{11}^2 \psi_0 = - \frac{1}{2 \pi\r} ( J_1 +J_2),&
\partial_{11}\psi_0 |_{x^{11} =0} =\frac{1}{2} J_1,&
\partial_{11}\psi_0 |_{x^{11}=\pi\r} =-\frac{1}{2} J_2 \nonumber \\
\partial_{11}^2 \psi_n = - \Box_4 \psi_{n-1},&
\partial_{11}\psi_n |_{x^{11} =0, \pi\r} =0,& n=1,2,... \nonumber \\
\label{eq:sol1}
\eea
and 
\bea
&< \psi_n> =0, & n=0,1,2,... 
\eea
This series has a recursive solution in terms of polynomials
$P_n, \; Q_n$, where
\be
 P_0 = - \frac{(x^{11})^2}{4 \pi\r} + \frac{x^{11}}{2} -
 \frac{\pi \r}{6}\; ,
 \quad Q_0 = - \frac{(x^{11})^2}{4 \pi\r} + \frac{\pi \r}{12} 
\label{eq:pzero}
\ee
and
\bea
 \partial_{11}^2 P_n=-P_{n-1}\; ,&\quad& \partial_{11}^2 Q_n=-Q_{n-1}\nn\\
 \left. \partial_{11} P_n\right|_{x^{11}=0}=0\; ,&\quad&
 \left. \partial_{11} Q_n\right|_{x^{11}=0}=0 \\
 <P_n> = 0\; ,&\quad&<Q_n> = 0\; ,
\eea
for $n>0$. For consistency, we have to check that the polynomials $P_n$,
$Q_n$ can really be chosen to fulfill all three conditions; that is, the
two boundary conditions and the vanishing average condition. This is,
\`a priori, not obvious since they are obtained by integrating a second
order differential equation and, therefore, contain only two free parameters.
Luckily, using the differential equation for $P_n$, the vanishing of the
average, $<P_{n-1}>=0$ implies that $P_n$ automatically satisfies the correct
boundary condition at $x^{11}=\pi\r$ (and the same for $Q_n$). The other two
conditions can be fulfilled by adjusting the two integration constants
so that $P_n$ is uniquely determined. It is easy to compute the successive
polynomials. For example,
\be
P_1 = \frac{(x^{11})^4}{48 \pi\r} - \frac{(x^{11})^3}{12}+
 \frac{\pi\r (x^{11})^2}{12} - \frac{\pi^3\r^3}{90}, \;
Q_1= \frac{(x^{11})^4}{48 \pi\r} -
 \frac{\pi\r (x^{11})^2}{24} + \frac{7 \pi^3\r^3}{720}
\label{eq:pol1}
\ee
Now, the question arises as to how quickly the series solution 
constructed above converges. As one can easily check, the ratio of
two successive terms in the series~\refs{series} is given by
$\psi_n/\psi_{n-1}=O\left( (\pi\r)^2/\s^2\right)$, where $\s$ is the scale
over which the gauge field varies (the instanton size). Formally, our
series provides a solution for any value of this ratio. For very small
instantons, however, the series might converge poorly.
If, on the other hand, the instanton size is sufficiently large as compared
to the separation $\pi\r$ of the boundaries, the series rapidly converges
and the solution for $\phi$ is well approximated by the first few terms
in the series.  Let us again consider the case of two instantons, one of
size $\sigma_1$ located at $r=0$ on the $x^{11}=0$ hyperplane, and the
other of size $\sigma_2$ located at $r=0$ at $x^{11}=  \pi \rho$.
If we assume that $\sigma_i 
\gg \pi \rho$ for $i=1,2$, then the solution for $\FO$ is well 
approximated by the first two terms in the series \refs{eq:psin}.
It follows from \refs{eq:curr},\refs{eq:psin},\refs{eq:pzero},\refs{eq:pol1}
that $\phi = \psi_0+\psi_1 +\cdots$ where
\be
\psi_0 = -48 \frac{\k^2}{\lambda^2} \left ( P_0 (x^{11}) \frac{\sigma_1^4}
{(\sigma_1^2 + r^2)^4}+  Q_0 (x^{11}) \frac{\sigma_2^4}
{(\sigma_2^2 + r^2)^4} \right ) 
\ee
and 
\be
\psi_1 = -768\; \frac{\k^2}{\lambda^2} \left ( P_1 (x^{11})
 \frac{3 r^2 -2 \sigma_1^2}
{(\sigma_1^2 + r^2)^6}+  Q_1 (x^{11}) \frac{3 r^2 -2 \sigma_2^2}
{(\sigma_2^2 + r^2)^6} \right ) \; .
\ee
We emphasize that this $x^{11}$-dependent solution 
represents a true strong coupling correction  to the gauge 5-brane. 

As an example, in Figure~1, we have plotted
$\phi$ as a function of $x=x^{11}$ and the 4--dimensional radius $r$
interpolating between two slightly different instantons located opposite
to each other on the two boundaries.
The separation of the boundaries has been chosen as $\r =1$ and the instantons
at $x^{11}=0$, $r=0$ and $x^{11}=\pi$, $r=0$ have the size $\s_1=11$ and $\s_2 =10$,
respectively. It was sufficient to use  the first two terms in the
series~\refs{series} only.

\epsfbox[-30 0 320 320]{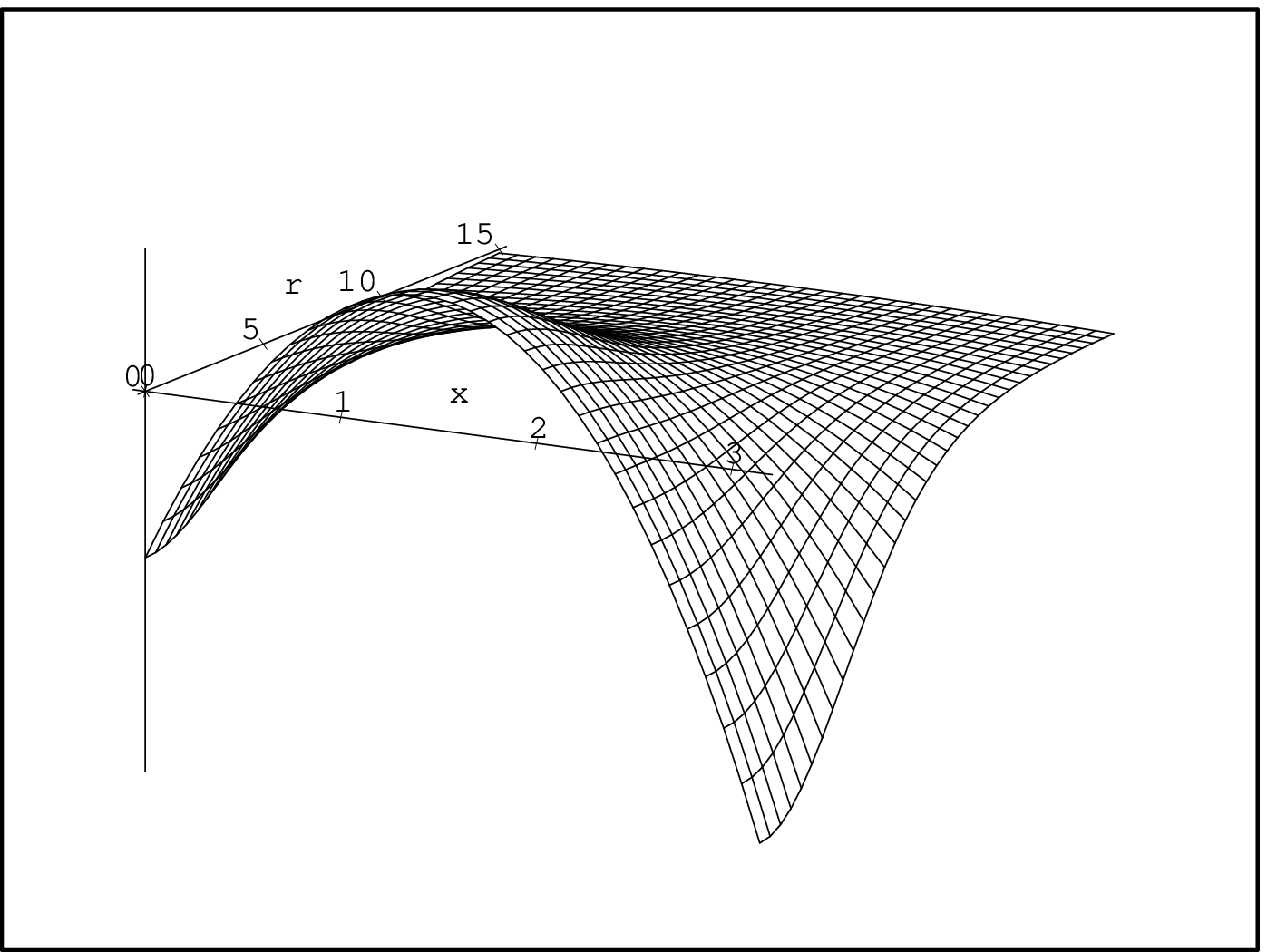}
\centerline{\em Fig 1: Correction $\phi$ to the string coupling
                       interpolating between two instantons.}
\vskip 0.4cm

We have not yet checked whether our solution preserves any supersymmetries.
This is, however, easily done since the vanishing of the 
supersymmetry variation
of the gravitino does not depend on the explicit solution for $e^C=\Phi$,
but rather  on the structure of the Ansatz and the relations between
$A$, $B$, $C$. Both are identical to the ones for the ordinary 5--brane.
Consequently, for spinors $\eta$ of the form
\be
 \eta =\e\otimes\r\; ,\quad \r = \r_0\;  e^{A/2}
\ee
and
\be
 (1\pm\g )\e = 0 \label{g5_chir}
\ee
the gravitino supersymmetry variation~\refs{susy} vanishes~\footnote{The
decomposition for spinors and gamma matrices is the same as for the ordinary
5--brane explained below eq.~\refs{5_chir}.}. Here, the chirality sign is
the same as the one in the Ansatz~\refs{5g_ans} for $G$, and is consequently
$+$ ($-$) for (anti)--selfdual instantons. Recall that, in the construction
of the solution, we have chosen the instantons on both boundaries to be of
the same type. Here we find that this is necessary to preserve any
supersymmetries. 
 Along with the $Z_2$ condition
$(1\pm\s_{11})\r_0 = 0$, eq.~\refs{g5_chir} implies that the solution
preserves $1/4$ of the 11-dimensional supersymmetry and $1/2$ of
the 10--dimensional supersymmetry of M--theory on $S^1/Z_2$.
If we had chosen instantons of different types, we would
find two chirality conditions of opposite sign from the two boundaries,
 thereby
projecting out the full spinor.

\vskip 0.4cm

Finally, we would like to discuss the precise relation of our
11--dimensional solution to the corresponding 10--dimensional gauge
5--brane. Generally, since our solution is the strongly coupled version
of the gauge 5--brane, we expect it to consist of a $x^{11}$-independent
bulk piece identical to the weakly coupled gauge 5--brane plus
$x^{11}$--dependent strong coupling corrections. The field strength
$H_{abc}$ of the 10--dimensional Neveu--Schwarz 2--form is the zero mode
of $G_{abc11}$, which can be computed by averaging 
\be
 H_{abc}\equiv \frac{1}{\sqrt{2}}<G_{abc11}>=
  \mp\frac{1}{2}{\hat{\e}_{abc}}^{~~~d}\partial_d<e^C>\; .
\ee
Let $\vf$ be the 10--dimensional dilaton related to the string coupling
$g_S$ by $g_S=e^{-\vf}$. Then, a comparison of the above equation with the
Ansatz for the weakly coupled gauge 5--brane leads us to identify
$g_S^2=e^{-2\vf}=<e^C>=\Phi_0$. Recall that $\Phi_0$ is the $x^{11}$--independent
part of our solution defined in eq.~\refs{ansa}. With this identification,
the Bianchi identity~\refs{B4} for $\Phi_0$ turns exactly into its
10--dimensional counterpart~\cite{strom}. More generally, a reduction of the
11--dimensional action to 10 dimensions on a metric with $g_{11,11}=e^{2B}$
leads to the identification~\cite{w2} $g_S^2=e^{-2\vf}=e^{3B}$. For our
solution $B=C/3$ so that $g_S^2=e^C$. Since the function $C$ depends on
$x^{11}$, this last equation is not quite correct, but
should be replaced by its averaged version $g_S^2=e^{-2\vf}=<e^C>=\Phi_0$.
This is in agreement with the previous result obtained by matching the
form fields. The relation between the 11--dimensional metric $g_{AB}$
and the 10--dimensional string frame metric ${}^{10}g_{AB}$ is given by
${}^{10}g_{AB}=e^{-2\vf /3}g_{AB}$. Our solution~\refs{g5_metr} for the
metric written in the 10--dimensional string frame then turns into
\be
ds^2_{10} = (1+e^{2\vf}\phi)^{-1/3}dx^\m dx^\n\eta_{\m\n}
            +e^{-2\vf}(1+e^{2\vf}\phi)^{2/3}dy^ady^b\d_{ab}\; .
\ee
Upon dropping the higher Fourier modes modes of $\phi$, this metric
coincides with the one for the weakly coupled gauge 5--brane~\cite{strom}.
In conclusion, we have seen that splitting up our solution as
$e^C=\Phi_0+\phi$, with an $x^{11}$--independent piece $\Phi_0$ and an
$x^{11}$--dependent correction $\phi$ with $<\phi>=0$, provides the
correct correspondence to the 10--dimensional gauge 5--brane. In particular
$\Phi_0$ equals the square of the string coupling.

\section{Conclusions}

In this paper, we have considered soliton solutions of M--theory on $S^1/Z_2$,
the low energy theory for the strongly coupled heterotic string. We have found
that the membrane solution of $D=11$ supergravity continues to be a solution
of M--theory on $S^1/Z_2$, including terms of relative order $\k^{2/3}$
for membranes oriented orthogonal, as well as parallel, to the $Z_2$
hyperplanes. Only in the first case, however, does the solution constitute a
BPS state of M--theory on $S^1/Z_2$. The term ``BPS'', in the present context,
is understood to label solutions which preserve $1/4$ of the 11--dimensional
supersymmetry and $1/2$ of the 10--dimensional supersymmetry (on the
$Z_2$ hyperplanes) of M--theory on $S^1/Z_2$. Membranes oriented parallel
to the hyperplanes, on the other hand, do not preserve any supersymmetries.
This reflects the fact that an orthogonal BPS membrane leads,
by dimensional reduction in the $x^{11}$--direction, to a BPS string,
 as
desired for the weakly coupled heterotic theory. A parallel BPS
membrane, on the other hand, would lead to a membrane of the weakly coupled
heterotic theory, which does not exist. It is interesting to trace the
nature of supersymmetry breaking for parallel membranes. The supersymmetry
variation of the gravitino can be set to zero near
any bulk point $x^{11}\in [-\pi\r ,\pi\r ]$ for appropriate spinors $\eta$.
Globally, however, we are faced with the sign flip in the chirality
condition~\refs{m_chir1} which projects out all globally defined spinors.
Hence, all supersymmetries are broken. This mechanism is reminiscent of
global supersymmetry breaking by gaugino condensation, discussed in
ref.~\cite{hor}.

The situation for the 11--dimensional 5--brane is similar,
 but with the
r\^ole of the orientations being reversed. For both possible orientations it
is a solution of M--theory on $S^1/Z_2$. However, the solution is BPS for
parallel 5--branes only. The orthogonal 5--brane, on the other hand, does
not preserve any supersymmetry. Again, this reflects the properties of the
weakly coupled heterotic theory in $D=10$, which has a 5--brane but not a
4--brane solution.

The 10-dimensional gauge 5--brane generalizes to the full 11--dimensional 
theory in a
nontrivial way. 
Unlike the membrane and 5-brane solutions, 
the gauge 5-brane does receive nontrivial 
corrections of order $\k^{2/3}$. This happens 
because the instantons on the hyperplanes switch on 
the anomalous terms in the Bianchi identity. We have presented 
a solution for this strongly coupled gauge 5-brane which makes 
its relation to the weakly coupled counterpart transparent. 
In particular, our solution contains an $x^{11}$-independent bulk component
which in a case of a single instanton exactly coincides with the weakly coupled gauge 5-brane. On top of this component comes an $x^{11}$-dependent 
part which is needed to interpolate between the instantons on different 
planes. It represents the strong coupling effect in the gauge 5-brane
solution. This $x^{11}$-dependent part has been computed in an 
expansion scheme which is quickly converging  as long as 
the gauge field varies slowly compared to the separation of the hyperplanes.
However, it should be noted that formally the solution we have given solves 
the equation for $\FO$ in general. In the present case it is possible to
approximate the solution very well just with the first two terms, 
but in more general cases the convergence may be much slower, in particular 
if one moves away from the hyperplanes.   
Clearly, this  method of finding solutions to the Horava-Witten model 
is not restricted to instanton type configurations, but can be applied to any 
physical gauge configuration on the boundaries. 
It is also particularly well suited to analyze the relation of the 11-dimensional theory to its 10-dimensional limit. 
An explicit example of the $x^{11}$-dependent part of our solution for two
 instantons of different sizes located opposite to each other on different 
hyperplanes has been depicted in Figure 1. 
For instantons of the same type (both selfdual or both anti-selfdual)
our solution preserves  one-quarter  of the 
11-dimensional supersymmetry and one-half of the 10-dimensional 
supersymmetry. 

These gauge 5-brane solutions are the first explicit examples of soliton solutions in the field-theoretical limit of M-theory on $S^1 / Z_2$ which receive 
nontrivial strong-coupling corrections. 
     
\vspace{0.4cm}

{\bf Acknowledgments} It is a pleasure to thank Dan Waldram for helpful
discussion. Z.~L.~and B.~A.~O.~are supported in part by the US--Polish Maria
Sklodowska Curie Joint Fund. A.~L.~is supported by Deutsche
Forschungsgemeinschaft (DFG) and Nato Collaborative Research Grant
CRG.~940784. B.~A.~O is partially supported by a Senior Alexander von
Humboldt Award.
\end{document}